\documentclass[aps,prb,longbibliography,showpacs,floatfix,superscriptaddress,twocolumn]{revtex4-1}

\usepackage{graphicx,color}
\usepackage{amsfonts}
\usepackage[figuresright]{rotating}
\usepackage{amssymb}
\usepackage{amsmath}
\usepackage{mathtools}
\usepackage{psfrag}
\usepackage{subfigure}
\usepackage{multirow}
\usepackage{tabularx}
\usepackage{textcomp}
\usepackage{units}
\usepackage{lipsum}
\usepackage{soul}
\usepackage{titlesec}
\usepackage{times}
\usepackage{hyperref}
\usepackage{enumitem}

\DeclareMathAlphabet\mathbfcal{OMS}{cmsy}{b}{n}

\titlespacing*{\section}
{0pt}{1ex plus .25ex}{1ex plus 1ex}
\titlespacing*{\subsection}
{0pt}{1ex plus .25ex}{1ex plus 1ex}
\titlespacing*{\subsubsection}
{0pt}{1ex plus .25ex}{1ex plus 1ex}

\titleformat{\section}
 {\fontsize{10}{15}\bfseries }
  {\thesection}
  {1em}
  {}

\titleformat{\subsection}
  {\bfseries }
  {\thesubsection}
  {2em}
  {}

\titleformat{\subsubsection}
  {\itshape}
  {\thesubsubsection}
  {2em}
  {}

\def\beq{\begin{eqnarray}}
\def\eeq{\end{eqnarray}}

\renewcommand{\v}[1]{\ensuremath{\mathbf{#1}}} 

\let\baraccent=\= 
\renewcommand{\=}[1]{\stackrel{#1}{=}} 
\newcommand{\bk}{\boldsymbol{k}} 
\newcommand{\mc}[1]{\mathcal{ #1}} 

\newcommand{\bl}[1]{\textcolor{black}{#1}} 

\makeatletter

\titleclass{\subsubsubsection}{straight}[\subsection]

\newcounter{subsubsubsection}[subsubsection]
\renewcommand\thesubsubsubsection{\thesubsubsection.\arabic{subsubsubsection}}

\titleformat{\subsubsubsection}
   {\normalfont\normalsize\itshape \centering}{\thesubsubsubsection}{1em}{}
\titlespacing*{\subsubsubsection}
{0pt}{1ex plus .3ex}{1ex plus .3ex}

\makeatletter
\renewcommand\paragraph{\@startsection{paragraph}{5}{\z@}%
  {3.25ex \@plus1ex \@minus.2ex}%
  {-1em}%
  {\normalfont\normalsize}}
\renewcommand\subparagraph{\@startsection{subparagraph}{6}{\parindent}%
  {3.25ex \@plus1ex \@minus .2ex}%
  {-1em}%
  {\normalfont\normalsize}}
\def\toclevel@subsubsubsection{4}
\def\toclevel@paragraph{5}
\def\toclevel@paragraph{6}
\def\l@subsubsubsection{\@dottedtocline{4}{7em}{4em}}
\def\l@paragraph{\@dottedtocline{5}{10em}{5em}}
\def\l@subparagraph{\@dottedtocline{6}{14em}{6em}}
\makeatother

\makeatletter
\def\maketitle{
\@author@finish
\title@column\titleblock@produce
\suppressfloats[t]}
\makeatother

\begin{document}
\title{Multiplicative topological phases}
\author{Ashley M.\ Cook$^*$}
\affiliation{Department of Physics, University of California, Berkeley,
California, 94720, USA}
\affiliation{Max Planck Institute for Chemical Physics of Solids, Nöthnitzer Strasse 40, 01187 Dresden, Germany}
\affiliation{Max Planck Institute for the Physics of Complex Systems, Nöthnitzer Strasse 38, 01187 Dresden, Germany}

\author{Joel E.\ Moore}
\affiliation{Department of Physics, University of California, Berkeley,
California, 94720, USA}
\affiliation{Materials Sciences Division, Lawrence Berkeley National Laboratory, Berkeley, California 94720 USA}

\begin{abstract}
Symmetry-protected topological phases of matter have challenged our understanding of condensed matter systems and harbour exotic phenomena promising to address major technological challenges. Considerable understanding of these phases of matter has been gained recently by considering additional protecting symmetries, different types of quasiparticles, and systems out of equilibrium. Here, we show that symmetries could be enforced not just on full Hamiltonians, but also on their components. We construct a large class of previously unidentified multiplicative topological phases of matter characterized by tensor product Hilbert spaces similar to the Fock space of multiple particles. To demonstrate our methods, we introduce multiplicative topological phases of matter based on the foundational Hopf and Chern insulator phases, the multiplicative Hopf and Chern insulators (MHI and MCI), respectively. The MHI shows the distinctive properties of
the parent phases as well as non-trivial topology of a child phase. We also comment on a similar structure in topological superconductors as these multiplicative phases are protected in part by particle-hole symmetry. The MCI phase realizes topologically-protected gapless states that do not extend from the valence bands to the conduction bands for open boundary conditions, which respect the symmetries protecting topological phase. The band connectivity discovered in the MCI could serve as a blueprint for potential multiplicative topology with exotic properties. \end{abstract}
\maketitle

\section{Introduction}
The search for novel phases of matter -- and particularly phases of matter beyond the Ginzburg-Landau paradigm, known as topological phases of matter -- is now a vast and influential topic in condensed matter physics~\cite{haldane1988,kane2005a, kane2005b,bernevig2006, konig766, maciejko2011,burkov2011, parameswaran2014, xubelopolski2015, Lu2015, lv2015, yan2017, kuroda2017}. The search has more recently focused primarily on considering an extended set of protecting symmetries~\cite{fu2011, hsieh2012, ando2015, Bradlyn2016, Benalcazar2017, Wieder2018}, and on realizing topology in systems that are non-electronic, driven, or coupled to an environment~\cite{khanikaev2012, susstrunk2015, lindner2011, cayssol2013, khemani2016, shen2018, gong2018}, as our understanding of electronic topology in equilibrium and in isolation was thought to be complete for effectively non-interacting systems.

\begin{figure}[t]
\centering
\includegraphics[width=0.3\textwidth]{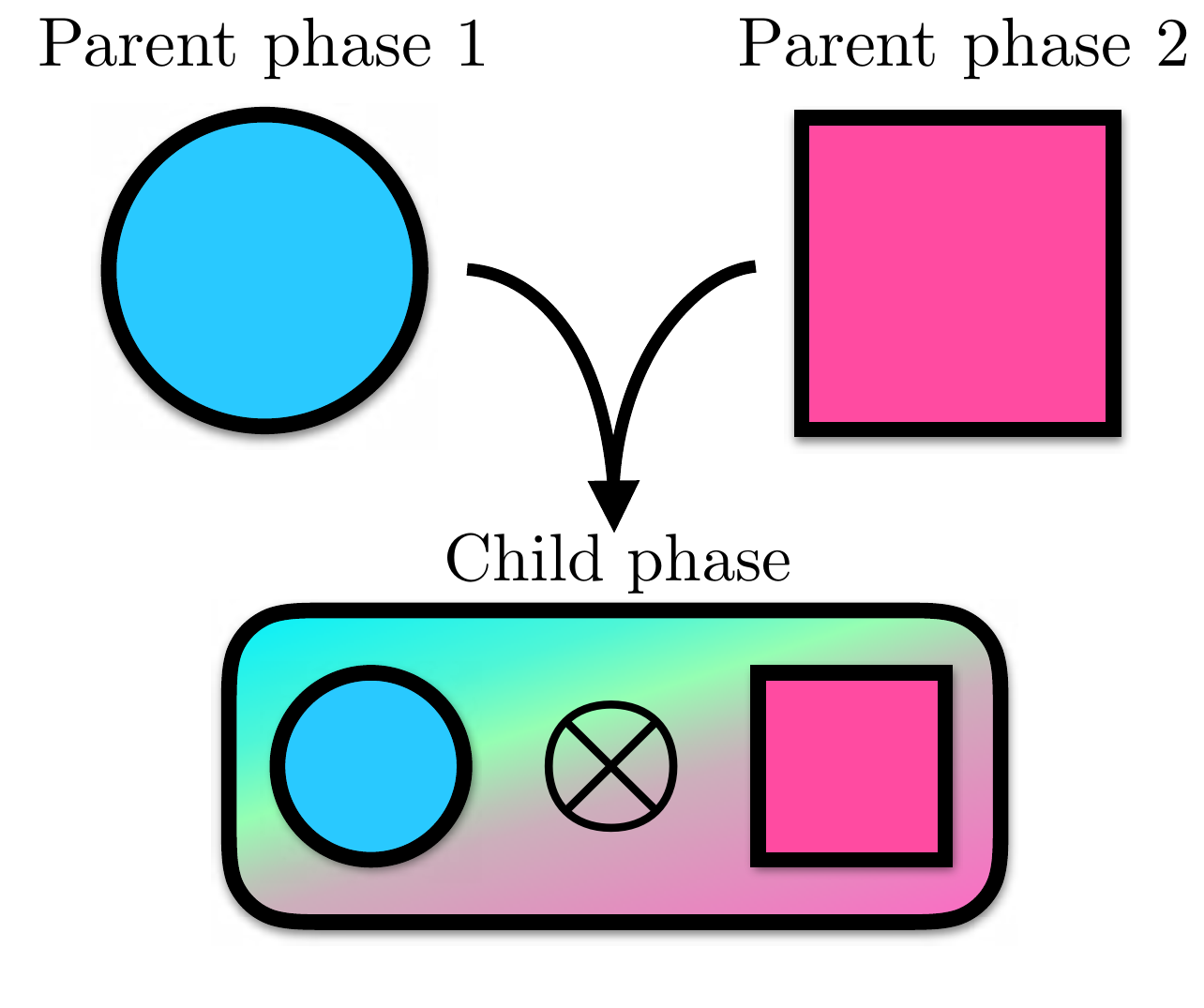}
\caption{
\textbf{Realization of multiplicative topology} Parent Hamiltonians $1$ and $2$, each possessing some symmetries and corresponding to a topological phase of matter, may be combined in a multiplicative manner to construct a single child Hamiltonian. This child Hamiltonian can possess an additional set of symmetries and inherit non-trivial topology from the parents in novel combinations. In this schematic, the first parent is represented by a blue circle, and the second parent is represented by a red square. The child is represented by a rectangle with rounded corners, with color interpolating between the blue of parent 1 and the red of parent 2, to symbolize inheritance by the child of features of each parent.}
\label{schematic}
\end{figure}

However, rather than realizing non-trivial topology by imposing symmetries only on the entire Hamiltonian for a system as has been done in the past, here we generalize by imposing symmetries on components of the Hamiltonian as well, which combine multiplicatively, i.e., via the vector space product rather than by the direct sum.  This approach leads to methods for construction of a large class of previously unidentified topological phases, including two examples of new phases as proof of concept of these methods along with an example of how a known phase can be viewed in this framework.  This approach considerably expands the possible set of symmetry-protected  topological phases of matter, as the large set of symmetries already considered in studying topology may be combined, allowing a set of parent phases of matter to be combined into a single child phase of matter synthesizing the properties of the parents, as illustrated in Fig.~\ref{schematic}. Many of these symmetries are commonplace, indicating multiplicative topology may appear in additional situations in electronic systems. This physics furthermore extends to various non-electronic topological phases and to driven or non-Hermitian systems, extending each of these sets of topological phases as well.

 We demonstrate this method explicitly by constructing two novel symmetry-protected topological phases of matter, the multiplicative Hopf insulator and the multiplicative Chern insulator, although the concept is broader; indeed we show that a product structure is naturally also present in one kind of topological superconductor. We begin by constructing an example of a multiplicative Hamiltonian that is the child of two non-degenerate parent Hamiltonians, $\mathcal{H}_1(\bk)$ and $\mathcal{H}_2(\bk)$. We then show that this form may be symmetry-protected such that it describes phases of matter, before introducing foundational examples of multiplicative phases of matter, the multiplicative Hopf and Chern insulators (MHI and MCI).

 \section{Results and Discussion}

 \subsection{Construction of multiplicative Hamiltonians}

 To construct the multiplicative Hamiltonian, we first determine the expressions for the matrix elements of the multiplicative Hamiltonian in terms of the matrix elements of the parent Hamiltonians. To do this, we consider parent Hamiltonians which are each acted upon by elements of the special unitary group, $\mathrm{SU}(2)$. Such Hamiltonians have two bands and may describe both the Chern insulator and the Hopf insulator topological phases of matter. A child Hamiltonian constructed from such parents would then be acted upon by the semisimple Lie group $\mathrm{SU}(2)\times\mathrm{SU}(2)$. As this Lie group is isomorphic to the double cover of the special orthogonal group $\mathrm{SO}(4)$, this direct product indicates there is a mapping from a pair of parent Hamiltonians, each with $2\times2$ matrix representation, to a child Hamiltonian with $4\times4$ matrix representation.

 The explicit construction of the isomorphism is given in the Methods, ``Expressing the multiplicative child Hamiltonian in terms of the parent Hamiltonian'', and here we state the resulting expression for the child Hamiltonian. We take $\mathcal{H}_1(\bk) = \begin{pmatrix} a & b \\ c & d\end{pmatrix}$ and $\mathcal{H}_2(\bk) = \begin{pmatrix} \alpha & \beta \\ \kappa & \delta \end{pmatrix}$ to be the two parent Hamiltonians in the construction, with momentum dependence suppressed. The expressions can of course be simplified further given hermiticity of $\mathcal{H}_1(\bk)$ and $\mathcal{H}_2(\bk)$, which gives $c = b^*$ and $\kappa = \beta^*$, ensuring hermiticity of the child Hamiltonian, but we leave the expressions more general to more clearly show the underlying dependence of the child Hamiltonian on the parent Hamiltonians.  We may then write the child Hamiltonian $\mathcal{H}_{c}(\bk)$ in terms of $\mathcal{H}_1(\bk)$ and $\mathcal{H}_2(\bk)$ as
\begin{equation}
\mathcal{H}_{c}(\bk) = \begin{pmatrix}
a \delta & -a \kappa & b \delta & -b \kappa \\
-a \beta & a \alpha & -b \beta & b \alpha \\
c \delta & -c \kappa & d \delta & -d \kappa \\
-c \beta & c \alpha & -d \beta & d \alpha
\end{pmatrix},
\end{equation}
Probably the most familiar physical examples of Hamiltonians of this product form, derived from underlying two-by-two Hamiltonians that can be expanded over Pauli matrices, appear in the theory of 2D Dirac materials, where the two-by-two components can represent spin, valley, or layer degrees of freedom. \bl{ More specifically, the additional particle-hole symmetry enforced to symmetry-protect the multiplicative phases in this work can correspond to the child effectively possessing a spinless version of time-reversal symmetry. The multiplicative phases could correspond, therefore, to a spinless superconductor with a suitable orbital or valley degree of freedom.}

\bl{We comment briefly on the differences between this tensor product construction of the multiplicative phases and square root topological phases~\cite{kane_2014, arkinstall2017, kremer_2020, Yoshida_2021}: these past works are distinct from our multiplicative constructions and involve finding a Bloch Hamiltonian $\mc{H}(\boldsymbol{k})$ with non-symmorphic symmetries at the expense of breaking crystal symmetries, starting from $\mc{H}^2(\boldsymbol{k})$. That is, these constructions are based on matrix multiplication rather than a tensor product, yielding phenomena which may be understood in the framework of crystalline topological phases. One example that highlights the difference is a two-dimensional multiplicative Bloch Hamiltonian $\mc{H}_c(k_x,k_y)$ can be constructed from two parents $\mc{H}_1(k_x)$ and $\mc{H}_2(k_y)$, which are each only one-dimensional and a function of momentum in the $x$-direction, $k_x$, or momentum in the $y$-direction, $k_y$. Such constructions are explored in detail in follow-up works currently being completed~\cite{Pal_Majorana2021, Pal_MWeyl2021}.}

\subsection{Topological classification}

The topological classification of the child Hamiltonian may be related to the homotopy classification of the parents as follows: each parent Hamiltonian can be diagonalized by an element of $\mathrm{SU}(2)$, but two diagonalizing elements are equivalent if they differ only by an element of the special diagonal unitary group with two elements $\mathrm{SDU}(2)$.\cite{avron1983}  We can obtain the Chern and Hopf invariants for each parent Hamiltonian by considering mappings from the $D$-dimensional Brillouin zone torus $\mathrm{T}^D$ to the space of distinguishable Hamiltonians, which is $SU(2) / U(1) = S^2$, where the $U(1)$ is the diagonal subgroup.\footnote{Strictly speaking the Hamiltonian in the ``periodic gauge'' is only periodic up to a gauge transformation; while this matters for some quantities such as polarization, it does not matter for the topological invariants we discuss.\cite{avron1983}}  An intuitive way to understand why the sphere appears here is to imagine expanding the Hamiltonian over Pauli matrices: the identity part is irrelevant (just a shifting of the zero of energy) and the non-degeneracy constraint means that the set of Hamiltonians is $\mathbb{R}^3$ with the origin removed, which has the same homotopy as the sphere $S^2$.

The corresponding homotopy group of the child Hamiltonian is similarly a mapping from a sphere to two spaces, each corresponding to an observable of a parent Hamiltonian. For $D=2$ or $D=3$, each parent Hamiltonian acted on by $\mathrm{SU}(2)$ corresponds to homotopy group $\pi_D (\mathrm{S}^2)=\mathbb{Z}$ for mappings to observables of the Hamiltonian. This gives topological classification of $\pi_D(\mathrm{S}^2) \times \pi_D(\mathrm{S}^2) = \mathbb{Z} \times \mathbb{Z}$ for the child Hamiltonian.  Mappings from the Brillouin zone torus are more complicated than mappings from the sphere, because the former do not necessarily form a group, but each nontrivial homotopy group $\pi_D$ of a space of Hamiltonians allows a new topological band structure starting in dimension $D$.  The $\pi_2$ invariant is known as the Chern number of a band and additional integer invariants appear for additional bands, while the $\pi_3$ Hopf invariant is more subtle and requires an additional generalized particle-hole symmetry $\mc{C}'$ to be defined in the case of additional bands~\cite{liu2017}.

\bl{It is possible for two topologically non-trivial phases of matter, each governed by its own non-trivial homotopy group, to co-exist, which could effectively yield $\mathbb{Z}\times \mathbb{Z}$ classification as one example. Our construction, however, instead allows properties of multiple topological phases to be integrated into one phase of matter, governed by a single homotopy group that is the direct product of two homotopy groups. Multiple topological phases can therefore be combined, rather than simply co-exist, such that the resultant child phase may inherit exotic combinations of phenomena of parent phases. Two one-dimensional topological phases can even be combined into a single two-dimensional phase, for instance. Such a two-dimensional phase is constructed from two one-dimensional Kitaev chain models in this way in a follow-up work~\cite{Pal_Majorana2021}.}

\subsection{Symmetry-protection}

Importantly, this multiplicative form for Hamiltonians can furthermore be symmetry-protected rather than being fine-tuned, such that it describes a phase of matter.
To identify protecting symmetries for the topological phases of the child Hamiltonians of interest to us, we proceed as follows: we change the form of the quotient to more easily relate the multiplicative Hamiltonian to Hamiltonians in different symmetry classes of the ten-fold way~\cite{schnyder2008}. We will then identify a close relationship between the multiplicative Hamiltonian and those Hamiltonians in class D and class DIII of the ten-fold way. Finally, we will show how we can protect the form of the multiplicative Hamiltonian using symmetries of classes D and DIII. This also reveals the general principle behind realization of symmetry-protected multiplicative topological phases of matter.

We first reformulate the quotient. To do this, we note that the child Hamiltonian may be acted upon by the double cover of $\mathrm{SO}(4)$ or $\mathrm{SO}(4)$ itself, and also possess observables that are $\mathrm{SO}(2)$ invariant, as $\mathrm{SO}(2)$ is isomorphic to $\mathrm{U}(1)$.

We next generalize this quotient to systems with more than four bands, which may more easily be connected to a particular set of protecting symmetries. To do this, we note that $\mathrm{SO}(4)/\mathrm{SO}(2)$ corresponds to the $N=2$ case of quotient $\mathrm{SO}(2N)/\mathrm{SO}(N)$.

We now compare this more general form for the quotient to the corresponding quotients of the classes of the ten-fold way in order to identify which symmetries a Hamiltonian must possess to yield observables acted on by $\mathrm{SO}(2N)/\mathrm{SO}(N)$, the general form of the quotient of the child Hamiltonian considered so far. To do this, we first note that Hamiltonians in class D of the ten-fold way possess observables acted upon by $\mathrm{SO}(2N)$, and those of class DIII are acted upon by $\mathrm{SO}(2N)/\mathrm{U}(N)$~\cite{schnyder2008}. This suggests symmetries corresponding to class D and symmetries corresponding to class DIII might be combined to realize the desired quotient.

To see how symmetries of classes D and DIII may be combined to symmetry-protect the multiplicative phases, we consider how these symmetries restrict the form of a Hamiltonian. We first consider the simpler case of class D and then that of class DIII.
In the case of this class of the ten-fold way, observables acted upon by $\mathrm{SO}(2N)$ correspond to Hamiltonians with particle-hole symmetry, with the particle-hole operator $\mc{C}$ squaring to $+1$. This results from the fact that the Bloch Hamiltonian, $\mc{H}$, must satisfy the following expression

\begin{align}
-\mc{C} \mc{H}^{\top} \mc{C} = \mc{H}
\end{align}

where $\mc{H}^{\top}$ is the transpose of $\mc{H}$.

To mathematically show how symmetries may be used to restrict the Hamiltonian to a multiplicative form, it is useful at this point to assume that our Hamiltonian may be adiabatically deformed to a flat-band counterpart that is topologically equivalent to our original dispersive Hamiltonian, provided that occupied bands remain occupied and vice versa. Considering Bloch Hamiltonians with $2N \times 2N$ matrix representation, with $N$ fully occupied bands and $N$ fully unoccupied bands, we adiabatically deform our Hamiltonian to a flat-band counterpart expressed as

\begin{align}
\mc{H}(\bk) = U^{}(\bk) I_{N,N} U^{\dagger}(\bk)
\end{align}

where here $I_{N,N} = \mathrm{diag}(I_N,-I_N)$ and $I_N$ is the $N \times N$ identity matrix, and $U^{}(\bk) \in \mathrm{SO}(2N)$. To restrict ourselves to a set of Bloch Hamiltonians that are topologically-equivalent (assuming the flat band limit assumption holds), we restrict $U(\bk)$ to $U(\bk) = \mathrm{diag}(U_+(\bk), U_-(\bk))$ that does not move $I_{N,N}$.

We now consider the symmetries of class DIII. This class possesses the same particle-hole symmetry as class D as well as time-reversal symmetry  and chiral symmetry. We could use these symmetries to now express $U_-(\bk)$ in terms of $U_+(\bk)$, effectively reducing the configuration space to the quotient $\mathrm{SO}(2N)/\mathrm{U}(N)$. Before doing this, however, we enforce an additional symmetry on $U(\bk)$ that yields the quotient of the multiplicative Hamiltonian instead. To do this, we note (suppressing momentum dependence) that $U = \mathrm{diag}(U_+, U_-)$ may be further restricted to the form
\begin{align}
U = \mathrm{diag}(U^{+}_{1}U^{+}_{2}, U^{-}_{1}U^{-}_{2},U^{+}_{1}U^{-}_{2},U^{-}_{1}U^{+}_{2}),
\end{align}
expressed in terms of $N \times N$ elements $U_{1} = \mathrm{diag}(U^{+}_{1}, U^{-}_{1})$ and $U_{2} = \mathrm{diag}(U^{+}_{2}, U^{-}_{2})$. These $U_{1}$ and $U_{2}$ may be thought of as the unitary matrices diagonalizing two separate $N \times N$ Bloch Hamiltonians, $\mc{H}_1(\bk)$ and $\mc{H}_2(\bk)$.

\begin{align}
\mc{H}_1(\bk) &= U^{}_1(\bk) I_{N/2,N/2} U^{\dagger}_1(\bk) \\
\mc{H}_2(\bk) &= U^{}_2(\bk) I_{N/2,N/2} U^{\dagger}_2(\bk)
\end{align}
where here $I_{N/2,N/2} = \mathrm{diag}(I_{N/2},-I_{N/2})$, corresponding to $\mc{H}_1(\bk)$ and $\mc{H}_2(\bk)$ being restricted to half filling, similarly to $\mc{H}(\bk)$. We see that generic elements that will not move $I_{N/2,N/2}$ in each case are $U_{1} = \mathrm{diag}(U^{+}_{1}, U^{-}_{1})$ and $U_{2} = \mathrm{diag}(U^{+}_{2}, U^{-}_{2})$, respectively. Here, we can also see that $U$ expressed in terms of $U_{i,\pm}$, where $i \in \{1,2 \}$, satisfies the half filling restriction on the full $2N \times 2N$ Hamiltonian.

To realize Hamiltonians with observables acted on by $\mathrm{SO}(2N)/\mathrm{SO}(N)$, we require that $\mc{H}_1(\bk)$ and $\mc{H}_2(\bk)$ each lie in class D. This restricts $U_1$ and $U_2$ to $\mathrm{SO}(N)$ rather than $\mathrm{U}(N)$. Thus, the desired multiplicative form is realized for Hamiltonians in class DIII with an additional particle-hole symmetry corresponding to class D.

While this method of construction here utilizes symmetries of class D and DIII, other symmetries could also be enforced in this manner to realize other topologically non-trivial phases of matter. This approach may also be generalized to constructions combining more than two parents, parents with different matrix dimensions, and construction of Hamiltonians describing $D$-dimensional topological phases from Hamiltonians for $d$-dimensional systems, where $d<D$.

As discussed in greater detail in the Methods, ``Stabilizing multiplicative topological phases up to closing of the bulk gap in systems with more than four bands'', although we discuss the case of symmetry-protection specifically in the four-band case above, this symmetry-protection also stabilizes multiplicative topological phases in systems with more than four bands up to closing of the bulk gap, similarly to results of Liu~\emph{et al.}~\cite{liu2017}, as the relevant homotopy groups can be nontrivial for $N>4$. These phases are therefore new stable topological phases rather than fragile topological phases.

\bl{We may also consider how the multiplicative phases can be extended by additionally considering crystalline point group symmetries. As one example, if a two-dimensional system in the $x-y$ plane were described by a Bloch Hamiltonian with projector space $SO(2N)/SO(N)$ ($N \ge 4$) and furthermore possessed a mirror symmetry $\mc{M}_z$ taking $z \rightarrow -z$, one could block-diagonalize the Hamiltonian by going to the basis in which $\mc{M}_z$ was diagonal, with each block possessing projector space $SO(4)/SO(2)$. One could then potentially realize a two-dimensional multiplicative phase in each $\mc{M}_z$ subsector. This is one approach to constructing a topological crystalline phase from the Chern insulator phase used previously~\cite{ueno2013}.}

   \begin{figure}[t]
\centering
\includegraphics[width=0.48\textwidth]{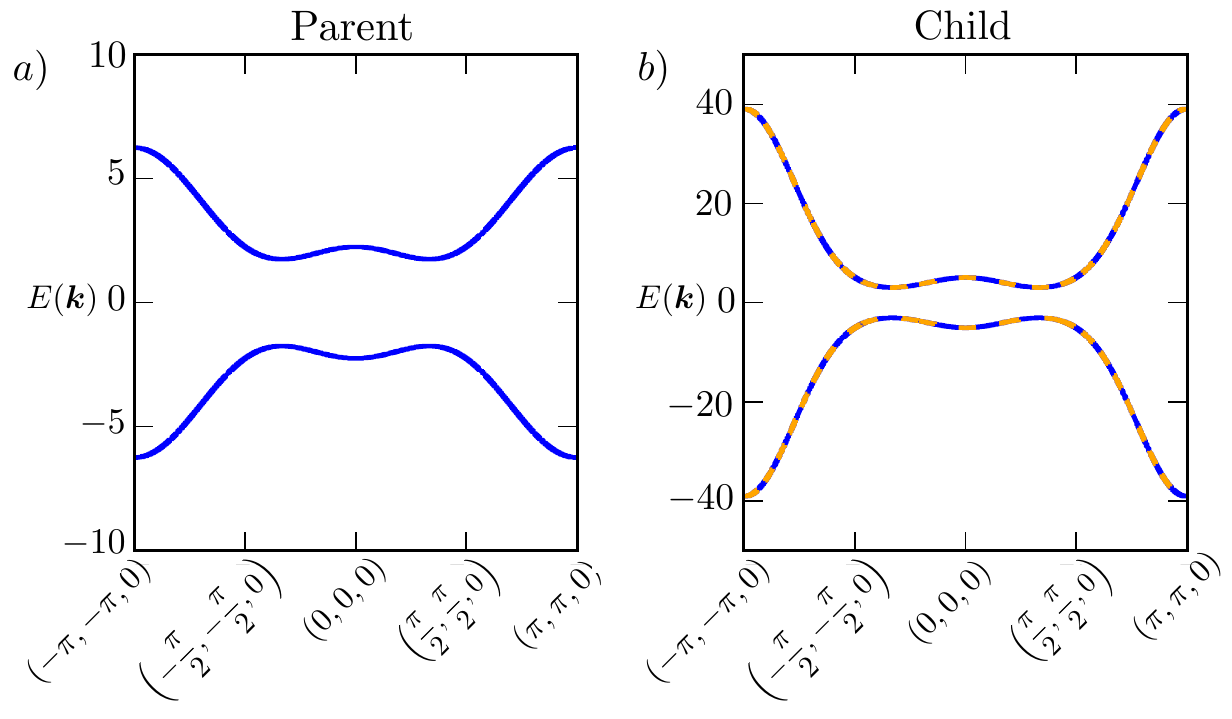}
\caption{\textbf{Comparison of parent bulk dispersion with child bulk dispersion.}
Bulk spectrum along a high-symmetry line for the Hopf insulator Hamiltonian, shown in (a) for $m=1.5$ as blue lines, and for the multiplicative Hopf insulator Hamiltonian, shown in (b), for $m=1.5$. The dashed blue and orange bands of the spectrum in (b) are doubly-degenerate.}
\label{fig1}
\end{figure}

\subsection{Multiplicative Hopf insulator}

To demonstrate our construction, we first combine two Hamiltonians, which describe the same $D$-dimensional topological phase, to realize a $D$-dimensional multiplicative phase. In particular, we choose a parent Hamiltonian with distinctive gapless boundary states in the topological phase to better compare the multiplicative topological phase with its parents. We also choose a parent phase of matter that has previously only been considered in isolation. For these reasons, we consider parent Hamiltonians characterizing the Hopf insulator phase. The Hopf insulator phase of matter has long been of interest as a topological phase realizable in non-interacting systems even without a protecting symmetry when the Hamiltonian is restricted to two bands. It has also recently gained prominence during the development of theory of fragile topological phases of matter~\cite{po2018}, given that its topology is unstable in the presence of additional bands without the addition of a protecting symmetry~\cite{liu2017}.

\begin{figure*}[t]
\centering
\includegraphics[width=1\textwidth]{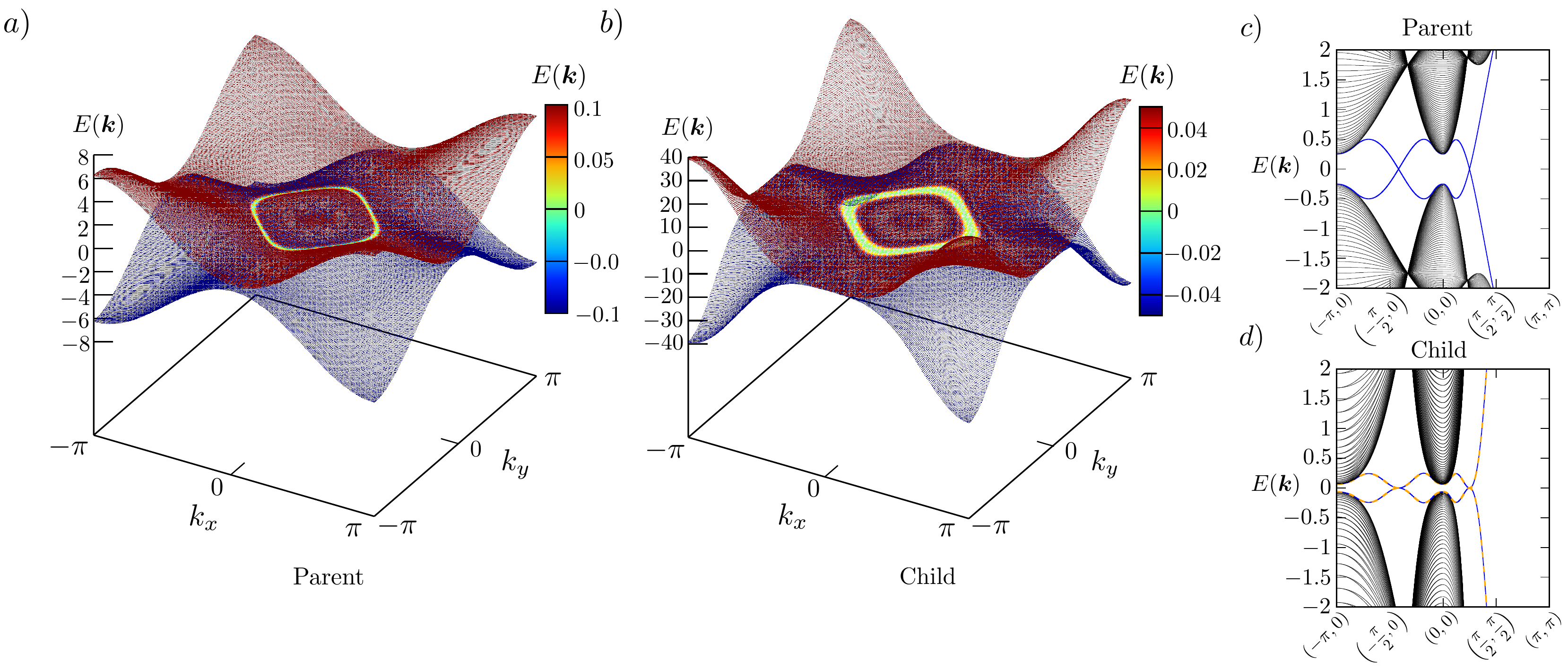}
\caption{\textbf{Comparison of bulk-boundary correspondence for parent topological phase with that of the child topological phase.} (a) The two middle bands of the surface spectrum for a Hopf insulator slab with open boundary conditions in the $\{001\}$ direction for $m=1.5$, showing the distinctive Fermi ring of topologically-protected gapless boundary states, and (b) the four middle bands of the surface spectrum for a multiplicative Hopf insulator slab with open boundary conditions in the $\{001\}$ direction for $m=1.5$, showing a double-degenerate Fermi ring of topologically-protected gapless boundary states. The number of layers in each slab calculations is $N = 80$, and the step size in $\bk$-space is $0.02$. The energy scales in (a) and (b) are high-lighted by color ranging from blue to red. (c) The full slab spectrum for the Hopf insulator along high-symmetry lines with open boundary conditions in the $\{001\}$ direction, with topologically-protected gapless boundary states of the two middle bands in energy highlighted in dark blue. (d) the full slab spectrum for the multiplicative Hopf insulator along high-symmetry lines with open boundary conditions in the $\{001\}$ direction, with topologically-protected gapless boundary states of the four middle bands highlighted as dashed blue and orange lines. Each dashed blue and orange line is doubly-degenerate. The number of layers in each slab calculations is $N = 80$, and the step size in $\bk$-space is $0.001$.
}
\label{fig3}
\end{figure*}

We will first show how to combine two Hopf insulator Hamiltonians into a single Hamiltonian characterizing the multiplicative Hopf insulator phase. We then characterize the multiplicative phase by studying its electronic structure for open boundary conditions. We will find that the multiplicative phase harbors topological gapless boundary modes quite similar to those of the Hopf insulator, both in appearance and in their robustness against surface perturbations.

 An established model for the Hopf insulator sufficient for our purposes here may be written as\cite{moore2008}
 \begin{equation}
 \mathcal{H}(\bk) = \boldsymbol{\nu} \cdot \boldsymbol{\sigma},
 \label{hopfeq}
 \end{equation}
 where $\nu^i = \boldsymbol{z}^{\dagger} \sigma^i \boldsymbol{z}$, $\boldsymbol{z} = \left( z_{\uparrow}, z_{\downarrow} \right)^{\top}$, $z_{\uparrow} = \sin k_x + i \sin k_y $, $z_{\downarrow} = \sin k_z + i\left(\cos k_x + \cos k_y + \cos k_z - m \right)$, and $\boldsymbol{\sigma}$ is the vector of Pauli matrices. Here, $k_x$, $k_y$, and $k_z$ are momentum in the $\hat{x}$-, $\hat{y}$-, and $\hat{z}$-directions, respectively, and $m$ is the single free parameter of the model that may be used to tune between the topologically trivial and non-trivial phases realized by the model.

Let us call the Hopf insulator in Eq.~\ref{hopfeq} $\mathcal{H}_{p1}(\bk)$. We may then also construct a time-reversed partner of $\mathcal{H}_{p1}(\bk)$, $\mathcal{H}_{p2}(\bk)$, using the time-reversal operator $\mathcal{T} = i \sigma_y \mathcal{K}$, where $\sigma_y$ is the second Pauli matrix and $\mathcal{K}$ is complex conjugation, as

 \begin{equation}
 \mathcal{H}_{p2}(\bk) = \mathcal{T} \mathcal{H}_{p1}(\bk)  \mathcal{T}^{-1}
 \end{equation}
 We construct a multiplicative Hopf insulator Hamiltonian $\mc{H}_{MH}(\bk)$ from the matrix elements of these two parent Hamiltonians, and first compare the bulk electronic structure of the Hopf insulator, shown in Fig.~\ref{fig1}(a), and the corresponding bulk electronic structure of the multiplicative Hopf insulator, shown in Fig.~\ref{fig1}(b). Along the high-symmetry line shown, there is a clear relationship between the eigenvalues of the multiplicative Hopf insulator and those of the Hopf insulator. Taking the lower-energy band of the electronic spectrum of the Hopf insulator to be  $\lambda_L(\bk)$, and the higher-energy band to be $\lambda_H(\bk)$, the lower-energy, doubly-degenerate bands of the multiplicative Hopf insulator spectrum have momentum-dependence $\gamma_L(\bk) = \mathrm{sgn}(\lambda_L(\bk))\lambda^2_L(\bk)$, and the higher-energy, doubly-degenerate bands of the multiplicative Hopf insulator spectrum have momentum-dependence $\gamma_H(\bk) = \mathrm{sgn}(\lambda_H(\bk))\lambda^2_H(\bk)$. This clearly reflects the product form of the multiplicative Hopf insulator Hamiltonian, with the two-fold degeneracy of eigenvalues further reflecting the protection of the phase by time-reversal symmetry.

To further explore the properties of the multiplicative Hopf insulator, we consider signatures of its topology in a slab calculation, with open boundary conditions in the $\hat{z}$-direction. An especially distinctive feature of the Hopf insulator is the Fermi ring of topologically-protected gapless boundary states on the surfaces of a slab with $N$ layers and open boundary conditions in the $\hat{z}$-direction, shown in Fig.~\ref{fig3} (a), bringing up the possibility that the multiplicative Hopf insulator should possess similar gapless boundary states. Indeed, we find a similar Fermi ring of topologically-protected states for the multiplicative Hopf insulator. Computing the slab spectrum for $\mathcal{H}_{MH}(\bk)$ for open boundary conditions in the $\{001\}$ direction, we find a four-fold degenerate Fermi ring of topologically-protected gapless boundary states for the appropriate value of $m$ of 1.5, shown in Fig.~\ref{fig3} (b). The bright green region highlighting the Fermi rings is wider here than in Fig.~\ref{fig3}(a): if we plot the slab spectra along high-symmetry lines in the surface Brillouin zone, as shown in Fig.~\ref{fig3}(d), we see this is due to the surface bands dispersing quadratically away from zero energy, while the bands disperse linearly from zero energy for the Hopf insulator, as shown in Fig.~\ref{fig3} (c). As expected and highlighted by the dashed blue and orange lines, we find the surface bands forming the Fermi rings are double-degenerate away from zero energy rather than non-degenerate as in the case of the Hopf insulator.

\begin{figure}[t]
\centering
\includegraphics[width=0.45\textwidth]{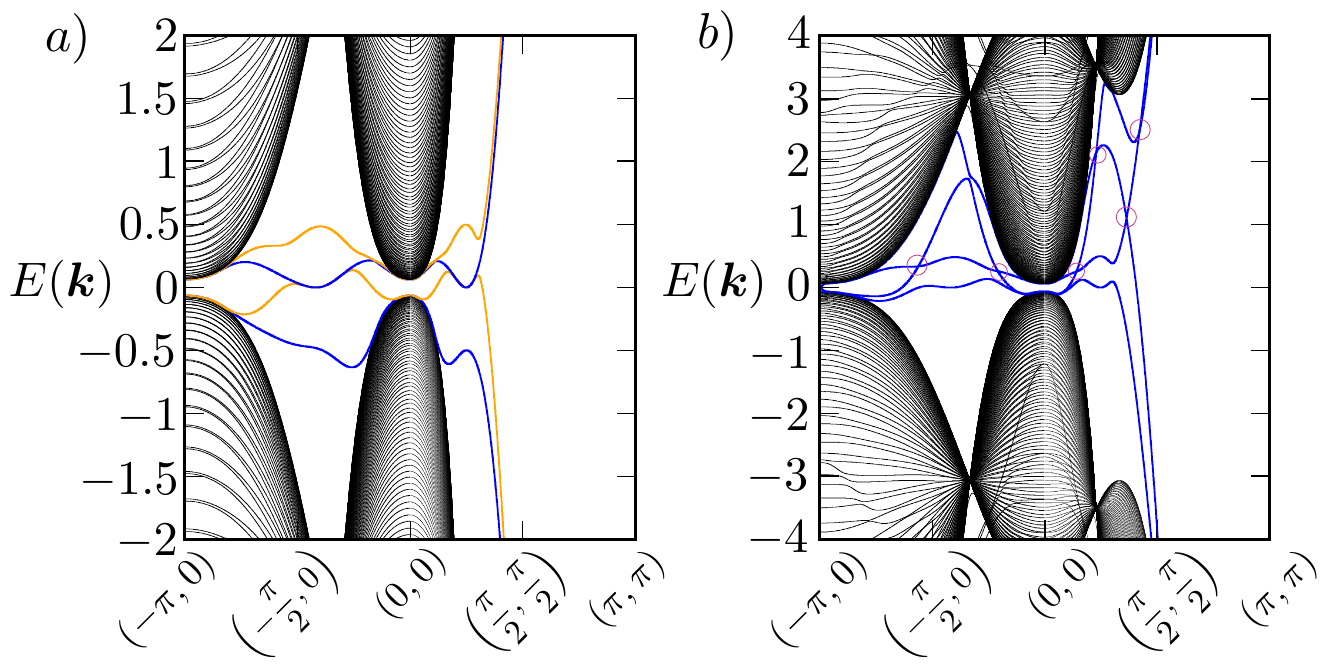}
\caption{\textbf{Effects of surface perturbations on the multiplicative Hopf insulator phase.}
(a) Slab spectrum for the multiplicative Hopf insulator along high-symmetry lines in the surface Brillouin zone with hard boundary perturbation term $\mathcal{H}_{\mu}$, where here $\mu = 0.5$ and $m = 1.5$. The four middle bands in energy that were two-fold degenerate for $\mu = 0$, highlighted by dashed orange and blue lines in Fig.~\ref{fig3} (d), are here highlighted alternatingly in blue and orange to emphasize the breaking of this two-fold degeneracy and the gapping out of one pair of the gapless boundary states to become topologically trivial, while the other pair remains topologically-protected, and (b) slab spectrum for the multiplicative Hopf insulator along high-symmetry lines in the surface Brillouin zone with soft, or adiabatic, boundary perturbation term $\mathcal{H}_{\alpha, \ell}$, with Zeeman field strength $\alpha = 0.05$ and number of slab layers over which the boundary perturbation increases $\ell = 4$, with $m  = 1.5$. The four middle bands in energy are highlighted in blue with two-fold degeneracies highlighted by pink circles. Each dispersion here is for a slab calculation with $N = 80$ layers and the step size in $\bk$-space is $0.001$.}
\label{fig4}
\end{figure}

We can further see how similar the behavior of the multiplicative Hopf insulator is to that of the parent Hopf insulators by considering additional perturbation terms near the surfaces of the slab Hamiltonian with open boundary conditions in the $\{001\}$ direction. First, we add hard boundary terms breaking time-reversal symmetry to the first and last layers. Taking the $4\times 4$ matrix representation of the terms on the diagonal of the slab Hamiltonian for layer $i$ to be $\mathcal{H}_{MH, i}$, this means taking the diagonal block for the $i=0$, or first layer, as $\mathcal{H}_{MH, 0} + \mathcal{H}_{\mu}$, and the diagonal block for the $i=N-1$ layer, or the last layer, as $ \mathcal{H}_{MH, N-1} + \mathcal{H}_{\mu}$. Here, we take the hard boundary perturbation term $\mathcal{H}_{\mu}$ to be a Zeeman field term $\mathrm{diag}\left(\mu, -\mu, \mu, -\mu \right)$. The slab dispersion for the multiplicative Hopf insulator along high-symmetry lines in the surface Brillouin zone with $\mu = 0.5$ is shown in Fig.~\ref{fig4} (a). Notably, in contrast to a conventional topological insulator protected by time-reversal symmetry, only some of the surface states are gapped out by the perturbation (in agreement with past work on the Hopf insulator~\cite{schuster2020}) rather than all of them, indicating these topologically-protected surface states are more robust than typically expected for a time-reversal invariant topological phase.

Instead of hard boundary perturbation terms, we may also consider soft, or adiabatic boundary perturbation terms. This involves taking $\mathcal{H}_{MH, i} \rightarrow \mathcal{H}_{MH, i} + \mathcal{H}_{i, \alpha, \ell}$, where $\mathcal{H}_{i, \alpha, \ell} = \mathrm{diag}\left(\alpha(\ell-i), -\alpha(\ell-i), \alpha(\ell-i), -\alpha(\ell-i) \right)$ for $i\le \ell$ and $\mathcal{H}_{i, \alpha, \ell} = \mathrm{diag}\left(\omega, -\omega, \omega, -\omega \right)$ for $i \ge N-1-\ell$ with $\omega = \alpha(N-1-i+\ell)$. $\mathcal{H}_{i, \alpha, \ell} = \mathrm{diag}\left(0, 0, 0, 0 \right)$ for other values of $i$. This corresponds to linearly-increasing perturbation terms near the boundaries of the slab. The physics of the soft boundary perturbation term is far richer than that of the hard term. As shown in Fig.~\ref{fig4} (b), it results in proliferation of topologically-protected degeneracies in the slab spectrum, where each such degeneracy is highlighted by a red circle. Such a boundary term has also previously been considered by others~\cite{moore2008, schuster2020, PhysRevA.85.063614} for the Hopf insulator and Chern insulator and has the same effect there as observed here, again indicating the multiplicative Hopf insulator behaves as expected, although a smaller value of $\ell$ is used here.  As discussed in past work~\cite{moore2008, schuster2020, schuster2019}, the degeneracies stem from the topology of the parent Hopf insulators: the \bl{two-fold} degeneracies are protected by translational invariance of the adiabatic edges that are smooth relative to the lattice length: as we still have translational invariance in the $\hat{z}$-direction because our edge is adiabatic, we may still think about a Bloch Hamiltonian defined over a three-dimensional Brillouin zone, now for each value of the adiabatically-varying parameter defining the adiabatic edge, and a two-fold degeneracy in a band-structure over a three-dimensional Brillouin zone, such as a Weyl node, does not require additional symmetry-protection, as can be seen from a simple $k \cdot p$ Hamiltonian for this case. (In the case of a three-dimensional topological insulator protected by time-reversal symmetry, in comparison, such adiabatic boundary conditions would yield gap-closings in the slab spectrum which are three-dimensional Dirac nodes, protected by translational invariance and time-reversal symmetry: if time-reversal symmetry were broken at this adiabatic boundary, one could lose these Dirac nodes as a gap-closing is then no longer required to transition from the strong topological insulator phase protected by time-reversal symmetry to the trivial vacuum.) The additional symmetries of the multiplicative topological phase are then just protecting the multiplicative structure stabilizing two Hopf insulators in a four-band system.

\subsection{Multiplicative Chern insulator}

We now present an example of a multiplicative topological phase, which exhibits signatures of non-trivial topology not realized in previously-known topological phases, to illustrate the potential for novel phenomena in multiplicative topological phases. For this reason, we choose to combine Hamiltonians for two Chern insulators into a single Hamiltonian modeling a multiplicative Chern insulator phase of matter. We will show the multiplicative Chern insulator phase realizes unusual band connectivity not observed previously to the best of our knowledge.

\begin{figure}[t]
\centering
\includegraphics[width=0.48\textwidth]{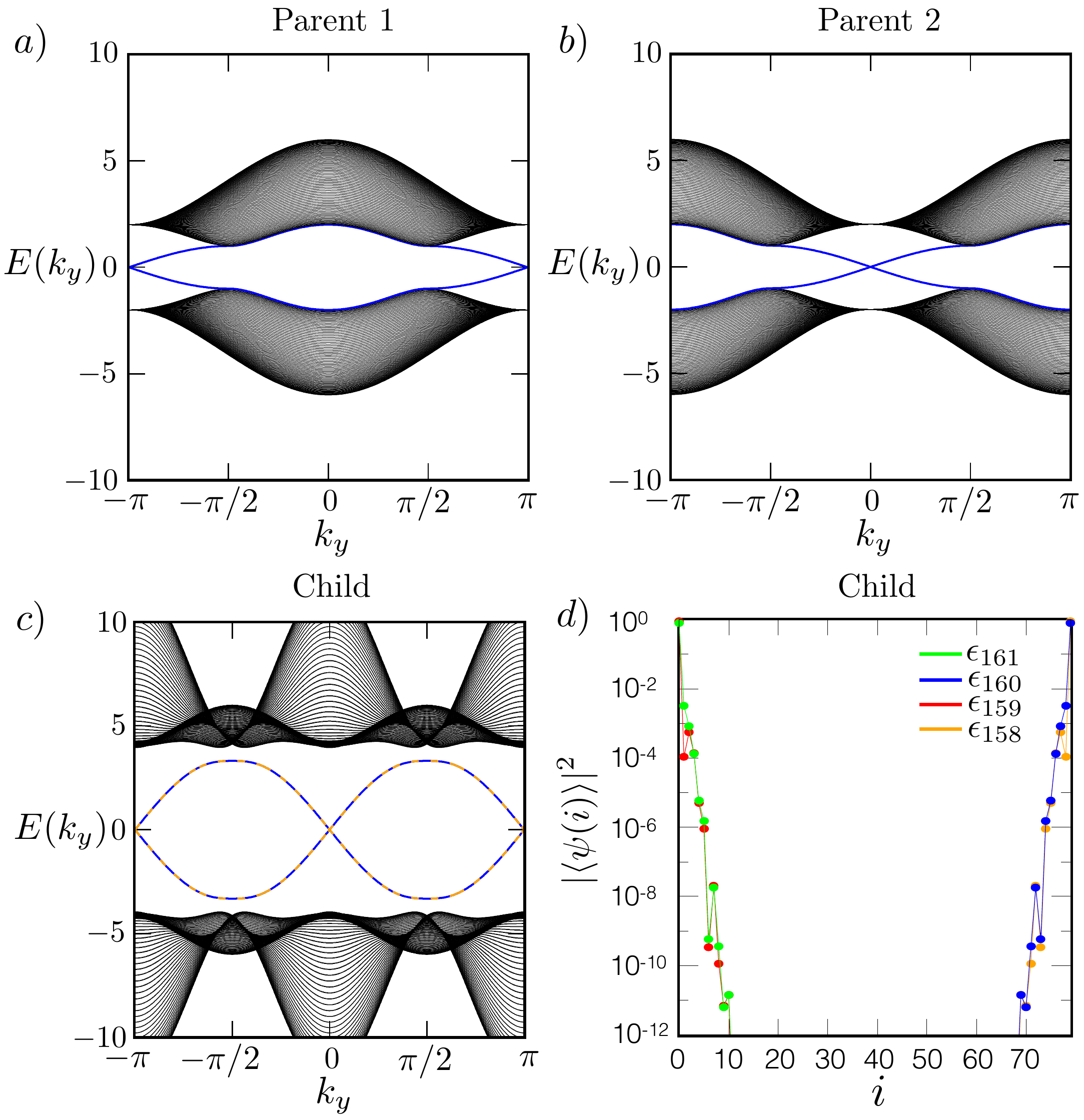}
\caption{\textbf{Emergence of novel phenomena in the child topological phase}
(a) Slab spectrum for the parent Chern insulator with $N=80$ layers and open boundary conditions in the $\hat{x}$-direction and periodic boundary conditions in the $\hat{y}$-direction, with free parameters $B$ and $M$ and $B=2$ and $M = -3$, showing linearly-dispersing topologically-protected gapless edge states highlighted in blue crossing at the edge of the slab Brillouin zone. (b) Slab spectrum for the parent Chern insulator with $N=80$ layers and open boundary conditions in the $\hat{x}$-direction and periodic boundary conditions in the $\hat{y}$-direction, with free parameters $B'$ and $M'$ and $B'=2$ and $M' = -1$, showing linearly-dispersing topologically-protected gapless edge states highlighted crossing at the center of the slab Brillouin zone. (c) Slab spectrum for the multiplicative Chern insulator constructed from the Chern insulators whose slab spectra are shown in (a) and (b), with $N=80$ layers and open boundary conditions in the $\hat{x}$-direction and periodic boundary conditions in the $\hat{y}$-direction. The two-fold degenerate bands corresponding to topologically-protected gapless edge states are highlighted as dashed orange and blue lines. The step size for (a), (b), and (c) in $\bk$-space is $0.01$. (d) Probability density $|\langle \psi (i) \rangle |^2$ vs. layer index $i$ for each of the four edge states computed for $k_y = \pi/4$ in the slab spectrum shown in (c), with the two states lower in energy labeled as $\epsilon_{158}$ (orange) and $\epsilon_{159}$ (red), and the two states higher in energy labeled as $\epsilon_{160}$ (blue) and $\epsilon_{161}$ (green), showing localization of each state at an edge of the slab. We also see that each edge possesses two counter-propagating edge states as expected for a time-reversal invariant system.}
\label{fig5}
\end{figure}

We take the well-known model for a Chern insulator on a square lattice~\cite{qi2006}, but with small adjustments. We write the model at a particular momentum $\bk = (k_x, k_y)$ as
\begin{equation}
\mathcal{H}(\bk) = \sin(k_x) \sigma_x + \sin(k_y) \sigma_y + \varepsilon(\bk) \sigma_z,
\end{equation}

where $\{\sigma_i\}$ are the three Pauli matrices and $\varepsilon(\bk) = B \left[2+M-\cos(k_x) - \cos(k_y)\right]$. Setting $B=1$, this Chern insulator has a topologically non-trivial phase with Chern number $C = +1$ for $-4 < M < -2$ corresponding to a pair of topologically-protected bands in the slab spectrum that are linearly-dispersive at low energy and cross at the boundary of the slab Brillouin zone, and a second topologically non-trivial phase with Chern number $C = -1$ for $-2 < M < 0$ corresponding to a pair of topologically-protected bands in the slab spectrum that are linearly-dispersive at low energy and cross at the center of the slab Brillouin zone.

Taking the first Chern insulator to have free parameters $B$ and $M$, we construct a second Chern insulator as the time-reversed partner of the first, but with free parameters $B'$ and $M'$, with $B'$ not necessarily equal in value to $B$ and $M'$ not necessarily equal in value to $M$. For $B = B'$ and $M = M'$, we can generate a multiplicative Chern insulator Hamiltonian $\mc{H}_{MC}(\bk)$.

We can also consider more interesting cases where $M' \neq M$, however (taking $B = B'$ for the cases considered here). Especially interesting is a case taking advantage of the difference in the location of the crossing of topologically-protected edge states depending on which topological phase is realized. When $M \in (-4, -2 )$ and $M' \in (-2,0)$ as in the case shown in Fig.~\ref{fig5}, the resulting topologically-protected gapless states of the MCI cross linearly at both the edge and the center of the slab Brillouin zone as shown in Fig.~\ref{fig5} (c), with two of the states localized on each edge as shown in Fig.~\ref{fig5} (d), clearly inheriting the crossings of the parent Chern insulators shown in Fig.~\ref{fig5} (a) and (b). Even more strikingly, however, these topologically-protected edge states do not extend into the masses of bulk valence and conduction bands. There is instead an energy gap separating the edge states from the bulk states, which does not scale with system size, controlled by $B$, as open boundary conditions produce deviations from the multiplicative form of the bulk near the boundary. As these boundary conditions \bl{do not break the symmetries protecting the topological phase}, this result serves as an example of novel band connectivity for gapped, non-interacting topological phases of matter, \bl{unlike in previous works, which disconnect gapless boundary states from the bulk valence and conduction bands \bl{in systems with trivial spectral flow in combination with breaking of symmetries protecting certain topological phases of matter}}~\cite{fengliu2017, potter2017, young2017}: it is consistent with standard band connectivity for the parent phases if the multiplicative structure exists in the bulk but not at the surface.

If one only examines band dispersions for single points in phase space, it is not clear how the disconnected boundary states of the MCI differ from disconnected boundary states due to breaking of a protecting symmetry at the boundary. Instead, to distinguish between the disconnected bands of the multiplicative topological phases and those of past works \bl{considering floating bands associated with trivial spectral flow}, it is important to consider how the disconnected states can be removed. In the case of the MCI, the disconnected states can only be removed by closing the bulk gap, and are therefore topologically-robust, while previously-studied disconnected states can be removed without closing the bulk gap, and are therefore not topologically-robust. Given the richness of topology protected in part by crystalline point group symmetries, we warn here that co-existing topological states---protected by different combinations of symmetries---confuse the issue of mechanisms for realizing floating boundary states in the literature: \bl{a topological state may co-exist with another trivial state ultimately yielding the floating boundary states, when a symmetry which could protect another topologically non-trivial state, such as spatial inversion symmetry in the case of weak stacks of SSH chains~\cite{fengliu2017, young2017, PhysRevB.102.161117}, is ill-defined due to open boundary conditions. (We discuss mechanisms for realizing such floating boundary states in greater detail in the Supplementary Note III, as the issue is subtle.)}

Concerning diagnostics for multiplicative topological phases, we note that the parity eigenvalues of the multiplicative phase are products of the parity eigenvalues of the parent phases. Thus band inversions could happen in the parent Hamiltonians, corresponding to change in sign of parent parity eigenvalues of occupied states, while still leaving the sign of the child parity eigenvalues unchanged, \bl{and we show this in the Supplementary Note I. We also note that Wilson loops may also fail to detect the non-trivial topology of multiplicative phases in the same situations, as the Wilson loop eigenvalues of the child are products of the Wilson loop eigenvalues of the parent phases (see Supplementary Note II).} This work therefore serves to initiate efforts to develop new methods that can identify these topological insulator phases overlooked by established approaches utilizing such band inversion diagnostics~\cite{tang2019, bradlyn2017}.

\section{Conclusions}

In conclusion, Hamiltonians describing symmetry-protected topological phases of matter may be combined to construct multiplicative topological phases of matter that exhibit properties of their parent phases as well as phenomena beyond our current understanding of non-trivial topology in materials. As any symmetries may be used, in principle, to protect multiplicative topological phases, they should be broadly realizable in materials and cold atom systems~\cite{RevModPhys.91.015005}. Similar physics may also be realizable in twisted bilayer graphene, however, given that models for this system can realize an approximate $\mathrm{SO}(4)$ symmetry~\cite{you_superconductivity_2019}. Counterparts of these topological phases are also expected in systems that are not purely electronic and/or not in equilibrium. Multiplicative topological phases can be realized by combining Hamiltonians with other symmetries beyond those discussed here, by combining more than two Hamiltonians, and also by combining Hamiltonians for lower-dimensional systems to form a higher-dimensional multiplicative phase. Given the exception to band connectivity discovered here in the case of the multiplicative Chern insulator, a foundational diagnostic of non-trivial topology in condensed matter systems, it is especially important to fully understand the phenomena that may result from this multiplicative topology.

\textbf{Acknowledgements} -  We thank E. Altman, I. Sodemann and R. Ilan for helpful discussions. A.~M.~C. also wishes to thank the Aspen Center for Physics, which is supported by National Science Foundation grant PHY-1066293, and the Kavli Institute for Theoretical Physics, which is supported by the National Science Foundation under Grant No. NSF PHY-1125915, for hosting during some stages of this work. A.~M.~C. was supported by the NSERC PDF and J.~E.~M. by NSF DMR-1918065.  Both authors acknowledge partial support from a Simons Investigatorship.

\textbf{Author contributions} - A.~M.~C. developed the concept, carried out the calculations, and led the writing of the manuscript.  J.~E.~M. advised on connections to previously studied topological structures and contributed to the writing.

\textbf{Data availability} - The data that support the findings of this study are available from the corresponding author upon request.

\textbf{Code availability} - The code that supports the findings of this study is available from the corresponding author upon request.

\textbf{Competing interests} - The authors declare no competing interests.

\textbf{Correspondence} - Correspondence and requests for materials should be addressed to A.M.C. (email: cooka@pks.mpg.de).

\bibliography{p1bib.bib}

\cleardoublepage

\section*{METHODS}

\subsection{Expressing the multiplicative child Hamiltonian in terms of the parent Hamiltonians}

We begin by constructing an example of a multiplicative Hamiltonian that is the child of two parent Hamiltonians, $\mathcal{H}_1(\bk)$ and $\mathcal{H}_2(\bk)$. We first determine the expressions for the matrix elements of the multiplicative Hamiltonian in terms of the matrix elements of the parent Hamiltonians. To do this, we consider action of an element of semisimple group $G = \mathrm{SU}(2) \times \mathrm{SU}(2)$, isomorphic to the double cover of $\mathrm{SO}(4)$, on an element in the space of special unitary $2 \times 2$ matrices,
\begin{equation}
\left(g_1,g_2 \right) \cdot g_3 := g_1 g_3 g_2^{-1},
\end{equation}
where $\left(g_1, g_2\right) \in G$ and $g_3$ is in the space of complex $2 \times 2$ matrices. We may also think of this action as quaternionic multiplication, an action which preserves vector length and which is linear in $g_3$~\cite{woit2017quantum}. This, in combination with $\left( g_1, g_2 \right)$ and $\left( -g_1, -g_2 \right)$ giving the same linear transformation of $\mathbb{R}^4$, reflects the fact that the action is an element of $\mathrm{Spin}(4)$.

We can write this action with respect to a particular basis $\{b_{11}, b_{12}, b_{21}, b_{22}\}$, where
\begin{eqnarray}
b_{11} &= \begin{pmatrix}
1 & 0 \\
0 & 0 \\
\end{pmatrix}
,
\hspace{2mm}
b_{12} = \begin{pmatrix}
0 & 1 \\
0 & 0 \\
\end{pmatrix}
, \\ \nonumber
\hspace{0mm}
b_{21} &= \begin{pmatrix}
0 & 0 \\
1 & 0 \\
\end{pmatrix}
,
\hspace{2mm}
b_{22} = \begin{pmatrix}
0 & 0 \\
0 & 1 \\
\end{pmatrix}.
\end{eqnarray}
We take $g_1 = \mathcal{H}_1(\bk) = \begin{pmatrix} a & b \\ c & d\end{pmatrix}$ and $g_2 = \mathcal{H}_2(\bk) = \begin{pmatrix} \alpha & \beta \\ \kappa & \delta \end{pmatrix}$, with momentum dependence suppressed. The expressions can of course be simplified further given hermiticity of $\mathcal{H}_1(\bk)$ and $\mathcal{H}_2(\bk)$, which gives $c = b^*$ and $\kappa = \beta^*$, ensuring hermiticity of the child Hamiltonian, but we leave the expressions more general for ease in following the construction. The action on each of the basis elements is
\begin{equation}
\bar{b}_{ij} = g^{}_1 b^{}_{ij} g^{-1}_2.
\end{equation}

We can express these actions as vectors instead of $2 \times 2$ matrices
\begin{eqnarray}
\bar{b}_{11} &= \left( a\delta , -a \beta , c \delta , -c \beta \right)^{\top}\\
\bar{b}_{12} &= \left(  -a \kappa , a \alpha , -c \kappa , c \alpha \right)^{\top} \\
\bar{b}_{21} &= \left(  b \delta ,  -b \beta , d \delta , -d \beta \right)^{\top} \\
\bar{b}_{22} &= \left(  -b \kappa ,  b \alpha , -d \kappa , d \alpha \right)^{\top}.
\end{eqnarray}
We can then represent these actions as a matrix $\mathcal{H}_{c} = \left( \bar{b}_{11}, \bar{b}_{12}, \bar{b}_{21}, \bar{b}_{22} \right)$, where
\begin{equation}
\mathcal{H}_{c}(\bk) = \begin{pmatrix}
a \delta & -a \kappa & b \delta & -b \kappa \\
-a \beta & a \alpha & -b \beta & b \alpha \\
c \delta & -c \kappa & d \delta & -d \kappa \\
-c \beta & c \alpha & -d \beta & d \alpha
\end{pmatrix},
\end{equation}
which is the matrix representation of the multiplicative Hopf insulator Hamiltonian in terms of the matrix elements of the Hopf insulator with matrix representation $g_1$ and its time-reversed partner with matrix representation $g_2$. While the Hopf insulator models considered, $\mathcal{H}_1(\bk)$ and $\mathcal{H}_2(\bk)$, include only nearest-neighbor hopping terms, the multiplicative Hopf insulator instead includes nearest- and next-nearest-neighbor hopping terms. The isomorphism between $\mathrm{Spin}(4)$ and $\mathrm{SU}(2) \times \mathrm{SU}(2)$ further guarantees an inverse mapping exists.

\subsection{Stabilizing multiplicative topological phases up to closing of the bulk gap in systems with more than four bands.}

Stabilizing the multiplicative phases in the presence of more than four bands may be done in the same way that the Hopf insulator is stabilized in the presence of more than two bands. In the case of the Hopf insulator as discussed in Liu~\emph{et al.}~\cite{liu2017}, the non-trivial homotopy group with only translational invariance and charge conservation for the $n+m$ band Bloch Hamiltonian with $n$ filled bands and $m$ unfilled bands is $\pi_3(\mathrm{Gr}(n,m+n))$. For $n=m=1$, this has $\mathbb{Z}$ classification, and trivial classification if one or both of $n$ and $m$ are greater than $1$. To stabilize the Hopf insulator for $n+m>2$, then, the homotopy group must change to some $\pi_3(M) \neq 0$: without a non-trivial homotopy group, a Hopf insulator could be adiabatically deformed to a topologically-trivial state. Then, in a system with more than two bands realizing a Hopf insulator phase, one must close the bulk gap to transition to a topologically-trivial phase.

This non-trivial homotopy group for more than two bands can be achieved through additional symmetry-protection. Importantly, the Hopf insulator Hamiltonian should be allowed by the additional symmetry constraints. One way to ensure this is by verifying that $\pi_3(M)$ reduces to $\pi_3(\mathrm{Gr}(1,2))$ for $n=m=1$.

For the Hopf insulator, a suitable $M$ is $\mathrm{Sp}(2n)/\mathrm{U}(n)$, as $\mathrm{Sp}(2)/\mathrm{U}(1) \cong \mathrm{Gr}(1,2)$, and $\pi_3(\mathrm{Sp}(2n)/\mathrm{U}(n)) \neq 0$. This homotopy group requires an additional generalized particle-hole symmetry $\mc{C}'$. With this constraint, the Hopf insulator phase is robust in systems with more than two bands up to closing of the bulk gap.

As we need some generalization of the quotient $\mathrm{SO}(4)/\mathrm{SO}(2)$ in the case of these first examples of multiplicative topological phases, there could be multiple avenues to symmetry-protection of the multiplicative topological phase in the presence of more than four bands. We discuss one option here that is similar to the case of the Hopf insulator. In this case, symmetries enforced at the level of the child Hamiltonian are time-reversal symmetry with $\mc{T}^2 = 1$ and spatial inversion symmetry $\mc{I}$.

This child then corresponds to a row of Table C2 of Ryu~\emph{et al.}~\cite{Ryu_2010}, with projector space $\mathrm{O}(n+m) / \left( \mathrm{O}(n) \times \mathrm{O}(m) \right)$. This is isomorphic to $\mathrm{O}(n+m) / \mathrm{O}(n) \times \mathrm{O}(n+m)/ \mathrm{O}(m) $. Enforcing an additional class D particle-hole symmetry on each parent, the projector space is reduced to $\mathrm{SO}(n+m) / \mathrm{SO}(n) \times \mathrm{SO}(n+m)/ \mathrm{SO}(m) $. This is isomorphic to $\mathrm{SO}(4)/\mathrm{SO}(2)$ for $n=m=2$.

We emphasize that, although spatial inversion symmetry is used in this example to stabilize the phase in the presence of more than four bands, the symmetry-protection we already considered also applies. For $n=m$, $\mathrm{O}(n+m) / \left(\mathrm{O}(n) \times\mathrm{O}(m)\right)$ is furthermore isomorphic to $\mathrm{O}(2n) / \mathrm{O}(n)$ and $\mathrm{SO}(2n) / \mathrm{SO}(n)$. Thus Table C2 also effectively gives us the homotopy groups for a child Hamiltonian in class DIII of the ten-fold way with additional class D symmetry enforced for each parent. With either of these combinations of symmetries, therefore, the four-band multiplicative phase is robust even in the presence of band mixing, up to closing of either parent bulk gap.







\end{document}